\documentclass[twocolumn,showpacs,amsmath,amssymb,prb,floatfix]{revtex4}
\usepackage{graphicx}% Include figure files
\usepackage{dcolumn}% Align table columns on decimal point
\usepackage{bm}% bold math
\begin{document}

\preprint{APS/123-QED}

\title{Signal propagation in time-dependent spin transport}

\author{Yao-Hui Zhu}
\author{Burkard Hillebrands}
\author{Hans Christian Schneider}
\email{hcsch@physik.uni-kl.de}
\affiliation{Physics Department and Research Center OPTIMAS,
  Kaiserslautern University of Technology, 67653 Kaiserslautern, Germany}


\date{\today}

\begin{abstract}
This paper analyzes theoretically the signal propagation in spin
transport by modulating the current passing through magnetic
multilayers. Using a macroscopic description of spin transport based
on the dynamical Boltzmann equation, we show that time-dependent
spin transport possesses a wave-like character that leads to
modifications of pure spin-diffusion dynamics. In particular, the
wave-like characteristics allow one to extract a finite spin
signal-propagation velocity.
%For long times, spin diffusion can
%be regarded as an approximation to the wave-diffusion duality of the
%time-dependent spin transport in the long-time limit.
\end{abstract}

\maketitle

\section{\label{sec:level1}Introduction}

Time-dependent spin transport in magnetic multilayers with current
perpendicular to the plane (CPP) is studied because of its
significance in physics and promising applications in spintronics
devices.~\cite{zut04,Fab07} Most theoretical investigations are based
on a diffusion equation for the spin accumulation or
magnetization.\cite{zhang02,Rash02,zhang05,cy06} These theories show
that if one drives a spin-polarized current through an interface from
a magnetic to a non-magnetic metallic layer, the spin propagates by
``diffusing'' into the nonmagnetic layers. If one considers
time-dependent spin transport, such as spin transfer torque
switching,~\cite{zhang02,zhang05} alternating current
(AC)~\cite{Rash02}, or magnetization switching,~\cite{cy06} where a
time-dependent signal is encoded in the spin orientation, one faces a
difficulty of the diffusion equation in that no propagation velocity
for the spin signal in the nonmagnetic layer can be defined. Or,
stated differently, the diffusion equation yields an infinite
propagation velocity for the spin signal in the metal, because the
signal will appear everywhere as soon as the source is switched
on.~\cite{zhang02} In this paper, we show how a physical propagation
velocity for spin signals in the CPP configuration can be determined
by deriving and analyzing macroscopic dynamical equations for
spin transport.

%% A problem analogous to that of an infinite signal propagation velocity
%% in the diffusion equation exists for the heat diffusion equation,
%% which gives an infinite velocity for the heat conduction. This
%% difficulty was resolved by substituting the Maxwell-Cattaneo
%% equation~\cite{Max67,Cat48} for Fourier's law, and deriving a
%% telegraph-type equation with a wave-diffusion duality for the heat
%% conduction.\cite{Joseph89,Scales99}

%% In the case of spin transport, we
%% show that the derivation of macroscopic equations, which yield a
%% finite propagation velocity, is possible by starting from the
%% Boltzmann equation instead of introducing directly an equation
%% analogous to the one used by Maxwell and Cattaneo.

It is an interesting connection that a problem analogous to that of an
infinite signal propagation velocity in the spin diffusion equation
exists for the heat diffusion equation, which yields an infinite
heat-conduction velocity. This difficulty was resolved by
recognizing that the theoretical description of heat transport needs
to be generalized by substituting the Maxwell-Cattaneo
equation~\cite{Max67,Cat48} for Fourier's law. In this way, one
obtains the physical picture that heat conduction is characterized by
a wave-diffusion duality. Formally, the heat diffusion equation needs
to be replaced by an equation that is essentially a telegraph
equation.~\cite{Joseph89,Scales99} As we show in this paper, a similar
modification of the spin-diffusion equation is necessary in the case
of spin transport.

We base our derivation of the macroscopic equations for spin transport
through multilayers on the theory developed for steady-state spin
transport across magnetic multilayers by Valet and Fert.~\cite{vf}
Instead of using the time-independent Boltzmann equation as in
Ref.~\onlinecite{vf}, we treat time-dependent spin transport starting
from the dynamical Boltzmann equation, which allows us to derive
macroscopic equations and to generalize the spin-diffusion equation.

This paper is organized as follows. The macroscopic dynamical
equations are derived in the Sec.~\ref{sec2} of our paper. Since the
central equations~(\ref{new7}) and (\ref{new8}), can also be cast in a
form that resembles telegraph equations, we discuss qualitative
aspects of dynamical spin transport in Sec.~\ref{sec3} using these
telegraph equations. In Sec.~\ref{sec5}, we analyze two concrete
examples of time-dependent spin transport numerically, and the main
conclusions are summarized in Sec.~\ref{sec6}.

\section{\label{sec2}Time-dependent equation system}

In this section, the model of Valet and Fert~\cite{vf} for
spin-dependent transport of conduction electrons through metallic
multilayers will be extended to take into account the time-dependence
of spin transport. The electron distribution function
$f_{s}(z,\mathbf{v},t)$ satisfies the linearized Boltzmann equation
\begin{equation}\label{bte}
\begin{split}
&\frac{\partial{f}_{s}(z,\mathbf{v},t)}{\partial{t}}+
v_{\rm{z}}\frac{\partial{f}_{s}(z,\mathbf{v},t)}{\partial{z}}
-eE(z,t)v_{\rm{z}}\frac{\partial{f}^{0}(v)}{\partial\varepsilon}\\
&=\!\int\!\!\!{d}^{3}v'\delta[\varepsilon(v')-\varepsilon(v)]P_{s}
[z,\varepsilon(v)]\big[f_{s}(z,\mathbf{v}',t)\!-\!f_{s}(z,\mathbf{v},t)\big]\\
&+\!\!\int\!\!\!{d}^{3}v'\delta[\varepsilon(v')\!-\!\varepsilon(v)]
P_{\rm{sf}}[z,\varepsilon(v)]\big[f_{-s}(z,\mathbf{v}',t)\!-
\!\!f_{s}(z,\mathbf{v},t)\big],
\end{split}
\end{equation}
where $-e$ and $\varepsilon(v)=mv^{2}/2$ denote, respectively, the
charge and kinetic energy of the electrons, and
$E(z,t)=-\partial{V}(z,t)/\partial{z}$ is the local electric
field.\cite{note1} $P_{s}(z,\varepsilon)$ and
$P_{\mathrm{sf}}(z,\varepsilon)$ are the spin conserving and
spin-flip transition probabilities, respectively. Following
Ref.~\onlinecite{vf}, we assume $f_{s}(z,\mathbf{v},t)$ to be the
sum of the Fermi-Dirac distribution $f^{0}(v)$ and small
perturbations:
\begin{equation}\label{appr}
f_{s}(z,\mathbf{v},t)\!=\!\!f^{0}(v)+\!\frac{\partial{f}^{0}}{\partial\varepsilon}\!
\left\{\left[\mu^{0}\!-\!\mu_{s}(z,t)\right]\!+\!g_{s}(z,\mathbf{v},t)\right\},
\end{equation}
where $\mu^{0}=mv_{\mathrm{F}}^{2}/2$ and $\mu_{s}(z,t)$ are the
equilibrium and nonequilibrium chemical potentials, respectively.
Due to the cylindrical symmetry of the system around the $z$ axis,
$g_{s}(z,\mathbf{v},t)$ can be expanded in Legendre polynomials of
$\cos\theta$, where $\theta$ is the angle between $\mathbf{v}$ and
the $z$ axis, as
\begin{equation}\label{lp}
g_{s}(z,\mathbf{v},t)=\sum_{n=1}^{\infty}g_{s}^{(n)}(z,t)P_{n}(\cos\theta).
\end{equation}
Here, the zero-order (isotropic) term is absent because
$(\partial{f}^{0}/\partial\varepsilon)g_{s}(z,\mathbf{v},t)$ was
defined by Eq.~(\ref{appr}) as the \textit{anisotropic} part of the
electron distribution perturbation. Using Eq.~(\ref{lp}), we obtain
\begin{equation}\label{rta}
\begin{split}
\frac{\partial{g}_{s}(z,\mathbf{v},t)}{\partial{t}}
+v_{\rm{z}}\frac{\partial{g}_{s}(z,\mathbf{v},t)}{\partial{z}}\!
+\!\left(\frac{1}{\tau_{s}}\!
+\!\frac{1}{\tau_{\rm{sf}}}\right)g_{s}(z,\mathbf{v},t)\\
=\frac{\partial\mu_{s}(z,t)}{\partial t}+v_{\rm{z}}\frac{\partial
\bar{\mu}_{s}(z,t)}{\partial{z}}
+\frac{\bar{\mu}_{s}(z,t)-\bar{\mu}_{\rm{-s}}(z,t)}{\tau_{\rm{sf}}},
\end{split}
\end{equation}
where $\bar{\mu}_{s}(z,t)=\mu_{s}(z,t)-eV(z,t)$ is the
electrochemical potential for electrons with spin $s$. The
derivation of this equation is detailed in Appendix \ref{derrta}.
Note that $\mathbf{v}$ in Eq.~(\ref{rta}) has been restricted to the
Fermi velocity $v_{\mathrm{F}}$, i.e.,
$\left|\mathbf{v}\right|=v_{\rm{F}}$ and
$v_{\rm{z}}=v_{\rm{F}}\cos\theta$.

With the relaxation times $\tau_{s}$ and $\tau_{\mathrm{sf}}$ (see
Eqs.~(\ref{rtas}) and (\ref{rtasf}) in Appendix \ref{derrta}), the
local electron mean free path $\lambda_{s}$, diffusion constant
$D_{s}$, and spin diffusion length $l_{s}$ can be defined,
respectively, as $\lambda_{s} =v_{\mathrm{F}}\tau'_{s}$,
$D_{s}=v_{\mathrm{F}}\lambda_{s}/3$, and
$l_{s}=(D_{s}\tau_{\mathrm{sf}})^{1/2}$, where the momentum relaxation
time $\tau'_{s}$ is defined by
\begin{equation}
1/\tau'_{s}=1/\tau_{s}+1/\tau_{\rm{sf}}.
\end{equation}
The appropriate ``average'' spin-diffusion length $l_{\mathrm{sf}}$
can be defined as
$(1/l_{\mathrm{sf}})^{2}=(1/l_{+})^{2}+(1/l_{-})^{2}$. Throughout
the paper, subscripts $+$ and $-$ stand for the \textit{absolute}
spin directions ``up'' and ``down'', respectively, whereas
subscripts $\uparrow$ and $\downarrow$ stand for the majority and
minority spin directions, respectively.

Using the method of Appendix B in Ref. \onlinecite{vf}, we
express the time-dependent current density for spin $s$ as
\begin{equation}\label{js}
J_{s}(z,t)=-\frac{e}{V}\sum_{\mathbf{v}}f_{s}(z,\mathbf{v},t)v_{\rm{z}}
=\kappa{g}_{s}^{(1)}(z,t),
\end{equation}
where $\kappa=\sigma_{s}/(e\lambda_{s})$. Note that $\kappa$ is
independent of $s$ and of the material in the Valet-Fert model. The
conductivity $\sigma_{s}$ can be written as
$\sigma_{s}=e^{2}n_{s}\tau'_{s}/m$, where
$n_{s}=4\pi(mv_{\mathrm{F}}/h)^{3}/3$ is the number of electrons
with spin $s$. It is easy to see that $\sigma_{s}$ satisfies
Einstein's relation $\sigma_{s}=e^{2}N_{s}D_{s}$, where
\begin{equation}
N_{s}=\frac{1}{4\pi^{2}}\left(2m/\hbar^{2}\right)^{3/2}\sqrt{\mu^{0}},
\end{equation}
is the density of states for spin $s$ at the Fermi level $\mu^{0}$,
and $N_{+}=N_{-}$.

Substituting Eqs.~(\ref{lp}) and (\ref{js}) into Eq.~(\ref{rta}), we
obtain
\begin{eqnarray}
\frac{e}{\sigma_{s}}\frac{\partial J_{s}(z,t)}{\partial
z}-\frac{1}{D_{s}}\frac{\partial \mu_{s}(z,t)}{\partial
t}=\frac{\bar{\mu}_{s}(z,t)-\bar{\mu}_{-s}(z,t)}{l_{s}^{2}},\quad\label{new1}\\
J_{s}(z,t)=\frac{\sigma_{s}}{e}\frac{\partial\bar{\mu}_{s}(z,t)}{\partial{z}}
-\tau'_{s}\frac{\partial {J}_{s}(z,t)}{\partial t}.\quad\label{new2}
\end{eqnarray}
In steady state, $J_{s}(z,t)$ and $\bar{\mu}_{s}(z,t)$ become
time-independent and then Eqs.~(\ref{new1}) and (\ref{new2}) reduce
to the Eqs.~(10) and (11) in Ref.~\onlinecite{vf}, respectively.

Equations (\ref{new1}) and (\ref{new2}) will be transformed to more
directly usable forms next. Without loss of generality, the
magnetization of the ferromagnet is set to be ``up''. Then, the
majority (minority) spins, which are antiparallel (parallel) to the
local magnetization (electron magnetic moment is
$\boldsymbol{\mu}=-(e/m)\mathbf{s}$) and denoted by subscript
$\uparrow$ ($\downarrow$), point to the absolute spin direction
``down'' (``up'') denoted by subscript $-$ ($+$). In terms of
$J_{\rm{m}}(z,t)=J_{+}(z,t)-J_{-}(z,t)$ and
$\mu_{\rm{m}}(z,t)=\mu_{+}(z,t)-\mu_{-}(z,t)$, Eqs.~(\ref{new1}) and
(\ref{new2}) can be written as
\begin{equation}
\frac{\partial{J}_{\rm{m}}(z,t)}{\partial{z}}
-eN_{s}\frac{\partial\mu_{\rm{m}}(z,t)}{\partial{t}}
=eN_{s}\frac{\mu_{\rm{m}}(z,t)}{T_{1}},\label{new3}
\end{equation}
\begin{equation}
J_{\rm{m}}(z,t)=eN_{s}\bar{D}
\frac{\partial\mu_{\rm{m}}(z,t)}{\partial{z}}
-\tau\frac{\partial{J}_{\rm{m}}(z,t)}{\partial{t}}
-\tilde{\beta}J(z,t),\label{new4}
\end{equation}
where
\begin{equation}\label{t1}
T_{1}=\tau_{\rm{sf}}/2
\end{equation}
can be regarded as the spin relaxation time.\cite{fl} The
``average'' diffusion constant $\bar{D}$ is defined as
$\bar{D}=c^{2}\tau$, with the wavefront velocity $c$ defined by
\begin{equation}\label{wfvelo}
c^{2}=v_{\rm{F}}^{2}/3.
\end{equation}
The ``average'' momentum relaxation time $\tau$ is
\begin{equation}\label{tau}
1/\tau=(1/\tau'_{+}+1/\tau'_{-})/2,
\end{equation}
and we have the identity $l_{\mathrm{sf}}=c\sqrt{\tau{T}_{1}}$. In
Eq.~(\ref{new4}),
$\tilde{\beta}=(\tau'_{-}-\tau'_{+})/(\tau'_{-}+\tau'_{+})$ equals
$\beta$ and $0$ for the ferromagnetic and nonmagnetic layers,
respectively. The bulk spin asymmetry coefficient $\beta$ in the
ferromagnetic layer is defined by
$\rho_{\uparrow(\downarrow)}=1/\sigma_{\uparrow(\downarrow)}
=2\rho_{\mathrm{F}}^{\ast}\left[1-(+)\beta\right]$, where
$\rho_{\mathrm{F}}^{\ast}$ is the total resistivity of the
ferromagnetic layer. In the nonmagnetic layer, we have
$\rho_{\uparrow(\downarrow)}=2\rho_{\mathrm{N}}^{\ast}$, where
$\rho_{\mathrm{N}}^{\ast}$ is the total resistivity of the
nonmagnetic layer.

In Eq.~(\ref{new4}), $J(z,t)$ stands for the total current density
$J(z,t)=J_{+}(z,t)+J_{-}(z,t)$. By introducing
$\mu(z,t)=\left[\mu_{+}(z,t)+\mu_{-}(z,t)\right]/2$, we can also
derive equations for the charge dynamics
\begin{equation}\label{charge1}
\frac{\partial{J(z,t)}}{\partial{z}}-
2eN_{s}\frac{\partial\mu(z,t)}{\partial{t}}=0,
\end{equation}
\begin{equation}
\begin{split}\label{charge2}
J(z,t)= \mbox{}&2eN_{s}\bar{D}\frac{\partial}{\partial{z}}\left[\mu(z,t)-eV(z,t)\right]\\
&-\tau\frac{\partial{J(z,t)}}{\partial{t}}-\tilde{\beta}J_{\mathrm{m}}(z,t).
\end{split}
\end{equation}

To describe spin accumulation by spin density instead of the
chemical potential, it is necessary to transform Eqs.~(\ref{new3})
and (\ref{new4}) using the following identity (Eq.~(\ref{nmu}) in
Appendix \ref{derden})
\begin{equation}
n_{\mathrm{m}}(z,t)=-eN_{s}\mu_{\mathrm{m}}(z,t),\label{den3}
\end{equation}
where $n_{\mathrm{m}}(z,t)=n_{+}(z,t)-n_{-}(z,t)$ is the spin
density and $n_{s}(z,t)$ the nonequilibrium charge density for spin
$s$. Using Eq.~(\ref{den3}), we can rewrite Eqs.~(\ref{new3}) and
(\ref{new4}) as
\begin{eqnarray}
\frac{\partial{J}_{\rm{m}}(z,t)}{\partial{z}}+\frac{\partial{n}_{\rm{m}}(z,t)}{\partial{t}}=
-\frac{n_{\rm{m}}(z,t)}{T_{1}},\label{new7}\\
J_{\rm{m}}(z,t)=-\bar{D}\frac{\partial{n}_{\rm{m}}(z,t)}{\partial{z}}
-\tau\frac{\partial{J}_{\rm{m}}(z,t)}{\partial{t}}
-\tilde{\beta}J(z,t).\label{new8}
\end{eqnarray}
To proceed further, we need the following identity (Eq.~(\ref{nmu2})
in Appendix \ref{derden})
\begin{equation}\label{total}
n(z,t)-2n_{s}^{0}=-eN_{s}\left[2\mu(z,t)-2\mu^{0}\right],
\end{equation}
where $n(z,t)=n_{+}(z,t)+n_{-}(z,t)$ is total nonequilibrium charge
density and $n_{s}^{0}$ the equilibrium charge density for spin $s$.
Using Eq.~(\ref{total}), we can rewrite Eqs.~(\ref{charge1}) and
(\ref{charge2}) as
\begin{equation}\label{charge3}
\frac{\partial{J(z,t)}}{\partial{z}}+\frac{\partial{n}(z,t)}{\partial{t}}=0,
\end{equation}
\begin{equation}
\begin{split}\label{charge4}
J(z,t)=&-\bar{D}\frac{\partial{n}(z,t)}{\partial{z}}-2e^{2}N_{s}\bar{D}
\frac{\partial{V}(z,t)}{\partial{z}}\\
&-\tau\frac{\partial{J(z,t)}}{\partial{t}}-\tilde{\beta}J_{\mathrm{m}}(z,t).
\end{split}
\end{equation}

In general, Eqs.~(\ref{new7}) and (\ref{new8}) should be solved
together with Eqs.~(\ref{charge3}), (\ref{charge4}), and Poisson's
equation. However, in metals and degenerate semiconductors, the
accumulation of charge occurs on a much smaller length scale and
varies much faster than that of
spin.~\cite{zhang02,Rash02,zhang05,cy06} Thus as an approximation,
it is assumed that the charge accumulation described by $n(z,t)$ can
always reach its steady state instantaneously when spin transport is
considered. This means that we always set
$\partial{n}(z,t)/\partial{t}=0$, which leads to
$\partial{J}(z,t)/\partial{z}=0$ according to Eq.~(\ref{charge3}).
Therefore, the current density $J(z,t)$ in Eqs.~(\ref{new4}) and
(\ref{new8}) becomes independent of $z$ and can be written as $J(t)$
instead.

\section{\label{sec3}``Telegraph'' equation}

In order to see the physical significance of the dynamics described
by Eqs.~(\ref{new7}) and (\ref{new8}) and to compare it with the
spin diffusion equation used in
Refs.~\onlinecite{zhang02,Rash02,zhang05,cy06}, we combine
Eqs.~(\ref{new7}) and (\ref{new8}) to yield the following equations
\begin{eqnarray}
\frac{\partial^{2}n_{\rm{m}}(z,t)}{\partial{t}^{2}}\!+\!(\frac{1}{\tau}
+\frac{1}{T_{1}})\frac{\partial{n}_{\rm{m}}(z,t)}{\partial{t}}
+\frac{n_{\rm{m}}(z,t)}{\tau{T}_{1}}\qquad\qquad\nonumber\\
=c^{2}\frac{\partial^{2}n_{\rm{m}}(z,t)}{\partial{z}^{2}},\label{new5}\quad
\end{eqnarray}
\begin{eqnarray}
\frac{\partial^{2}{J}_{\rm{m}}(z,t)}{\partial{t}^{2}}+(\frac{1}{\tau}+
\frac{1}{T_{1}})\frac{\partial{J}_{\rm{m}}(z,t)}{\partial{t}}
+\frac{J_{\rm{m}}(z,t)}{\tau{T}_{1}}\qquad\quad\nonumber\\
=c^{2}\frac{\partial^{2}{J}_{\rm{m}}(z,t)}{\partial{z}^{2}}-
\tilde{\beta}\left[\frac{1}{\tau}\frac{\partial{J}(t)}{\partial{t}}+
\frac{J(t)}{\tau{T}_{1}}\right].\label{new6}\quad
\end{eqnarray}
Because of the formal similarity of each of Eqs.~(\ref{new5}) and
(\ref{new6}) with the telegraph equation, we will refer to them as
telegraph equations in the following.

Each of the telegraph equations contains a second-order time
derivative, which is absent in the spin diffusion equation. This
term originates from the time derivative of the spin current in
Eq.~(\ref{new8}), which is also absent in the corresponding equation
for the spin current in spin diffusion theory; see, for instance,
Eq.~(8) of Ref.~\onlinecite{zhang02}. This additional term shows
that it takes a finite time for the spin current to adjust to the
gradient of the spin accumulation.~\cite{Joseph89,Che63} The
second-order time and space derivatives lead to a wave character of
dynamical spin transport in addition to its diffusion character
described by the first-order time and second-order space
derivatives. Thus, these equations show that time-dependent spin
transport should be understood using a wave-diffusion duality
picture. The occurrence of \textit{spin accumulation waves} enables
one to determine a well-defined propagation velocity $c$ for the
signal in time-dependent spin transport. Although the spin diffusion
equation does not yield spin accumulation waves and thus a finite
signal propagation velocity, it can be regarded as an approximation
of the wave-diffusion duality of the time-dependent spin transport
in the long-time limit.

\begin{figure}
\includegraphics[width=.45\textwidth]{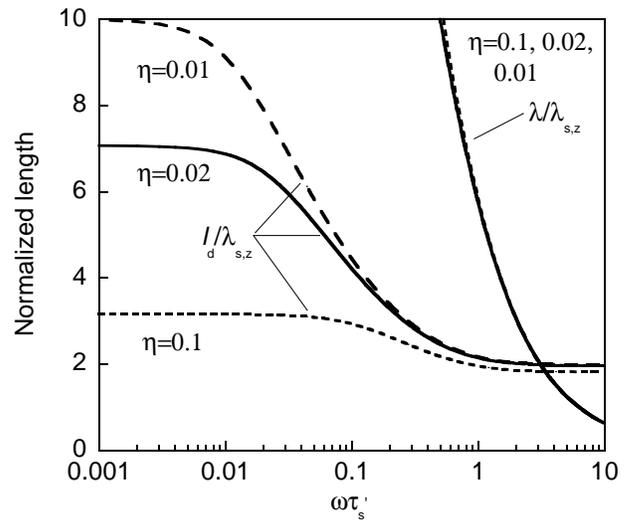}
\caption{Variation of $l_{\mathrm{d}}/\lambda_{s,\mathrm{z}}$
and $\lambda/\lambda_{s,\mathrm{z}}$ with $\omega\tau'_{s}$ for
three different values of $\eta$. The short-dashed, solid, and
long-dashed curves correspond to $\eta=0.1$, $0.02$, and $0.01$,
respectively. \label{fig1}}
\end{figure}

In the following, the telegraph equation of the nonmagnetic layer
will be analyzed in detail. Here, we have $\tau=\tau'_{s}$ and
$\tilde{\beta}=0$. Thus, Eqs.~(\ref{new5}) and (\ref{new6}) have the
same structure and we discuss only Eq.~(\ref{new5}) without loss of
generality. We seek a damped and dispersive wave solution to
Eq.~(\ref{new5}) of the form
\begin{equation}\label{solution}
n_{\mathrm{m}}(z,t)=n_{\mathrm{m}}^{0}\exp[i(kz-\omega{t})].
\end{equation}
At this stage, we can set either
$\omega=\omega_{\mathrm{r}}+i\omega_{\mathrm{i}}$ or
$k=k_{\mathrm{r}}+ik_{\mathrm{i}}$. The complex $\omega$ and $k$
will yield damping factors in time and space, respectively. Since we
are more interested in the damping length (or the dynamical spin
diffusion length), we will follow the method of
Ref.~\onlinecite{Kadin80} and assume
$k=k_{\mathrm{r}}+ik_{\mathrm{i}}$. Substituting
Eq.~(\ref{solution}) into Eq.~(\ref{new5}), we get the dispersion
relation
\begin{equation}\label{dispersion}
-\omega^{2}-i\alpha\omega+\xi=-c^{2}k^{2},
\end{equation}
where $\alpha=1/\tau'_{s}+1/T_{1}$ and $\xi=1/(\tau'_{s}T_{1})$.
Separating the real and imaginary parts of Eq.~(\ref{dispersion}),
we obtain
\begin{equation}
k_{\mathrm{r},\mathrm{i}}^{2}=\frac{1}{2c^{2}}\left[\sqrt{(\omega^{2}-\xi)^{2}
+\alpha^{2}\omega^{2}}\pm(\omega^{2}-\xi)\right],\label{kri}
\end{equation}

The wavelength, defined as $2\pi/|k_{\mathrm{r}}|$, can be written
as
\begin{equation}\label{wl}
\frac{\lambda}{\lambda_{s,\mathrm{z}}}=2\pi\sqrt{2}\left[
\sqrt{(\tilde{\omega}^{2}-\eta)^{2}+(\eta+1)^{2}\tilde{\omega}^{2}}
+(\tilde{\omega}^{2}-\eta)\right]^{-\frac{1}{2}},
\end{equation}
where $\lambda_{s,\mathrm{z}}=c\tau'_{s}$ is the $z$ component of
the electron mean free path. Moreover, we have introduced
dimensionless quantities
\begin{equation}
\tilde{\omega}=\omega\tau'_{s} \quad \text{and} \quad
\eta=\tau'_{s}/T_{1}.
\end{equation}

The damping length, defined as $l_{\mathrm{d}}=1/|k_{\mathrm{i}}|$,
can be written as
\begin{equation}\label{dl}
\frac{l_{\mathrm{d}}}{\lambda_{s,\mathrm{z}}}=\sqrt{2}
\left[\sqrt{(\tilde{\omega}^{2}-\eta)^{2}
+(\eta+1)^{2}\tilde{\omega}^{2}}
-(\tilde{\omega}^{2}-\eta)\right]^{-\frac{1}{2}}.
\end{equation}
Note that $l_{\mathrm{d}}$ can also be regarded as the
dynamical spin diffusion length.
When $\tilde{\omega}\rightarrow0$ or $\infty$, the damping length
$l_{\mathrm{d}}$ will approach $l_{\mathrm{sf}}$ or
$2l_{\mathrm{sf}}\sqrt{\tau'_{s}T_{1}}/(\tau'_{s}+T_{1})$,
respectively.

Figure~\ref{fig1} shows the variation of
$l_{\mathrm{d}}/\lambda_{s,\mathrm{z}}$ and
$\lambda/\lambda_{s,\mathrm{z}}$ with $\omega\tau'_{s}$ for three
different values of $\eta$. Note that the curves of
$\lambda/\lambda_{s,\mathrm{z}}$ for different $\eta$ are very close
to each other in the frequency range shown in the figure. The damping
length $l_{\mathrm{d}}$ decreases with frequency, which is
analogous to the skin effect of the electromagnetic wave propagating
in metal. The intersection of $l_{\mathrm{d}}/\lambda_{s,\mathrm{z}}$
and $\lambda/\lambda_{s,\mathrm{z}}$ indicates the critical angular
frequency $\omega_{\mathrm{c}}$, which separates the wave-like region
from the diffusion dominated regime, because the wave character
becomes significant only if the damping length exceeds the wavelength.
Stated differently, the wave character is significant if the typical
time scale $\tau_{\mathrm{sig}}$ of the time-dependent process is
smaller than the critical period
$T_{\mathrm{c}}=2\pi/\omega_{\mathrm{c}}$. On the contrary, the
diffusion character is dominant if
$\tau_{\mathrm{sig}}>{T}_{\mathrm{c}}$, and the spin-diffusion picture
becomes a good approximation of the wave-diffusion duality in the
limit $\tau_{\mathrm{sig}}\gg{T}_{\mathrm{c}}$.

An explicit expression for the critical angular frequency
$\omega_{\mathrm{c}}$ is obtained by combining
$\lambda=l_{\mathrm{d}}$ with Eq.~(\ref{kri}) 
\begin{equation}\label{critical2}
\begin{split}
\omega_{\mathrm{c}}\tau'_{s}&=\frac{1}{2}\left[
\gamma(1+\eta)+\sqrt{\gamma^{2}(1+\eta)^{2}+4\eta}\right]\\
&\approx\gamma+(\gamma+\frac{1}{\gamma})\eta,
\end{split}
\end{equation}
where $\gamma=\pi-1/(4\pi)\approx3.06$. Then, we have
$\omega_{\mathrm{c}}\tau'_{s}=3.06+3.4\eta$ approximately.

\begin{figure}
\includegraphics[width=.45\textwidth]{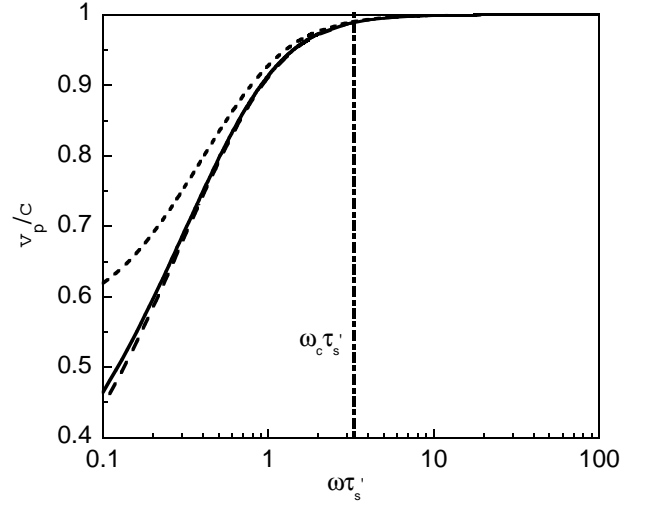}
\caption{The variation of $v_{\mathrm{p}}/c$ with $\omega\tau'_{s}$
for three different values of $\eta$. The short-dashed, solid, and
long-dashed curves correspond to $\eta=0.1$, $0.02$, and $0.01$,
respectively. The thick vertical dot-dashed line indicates the
critical angular frequencies $\omega_{\mathrm{c}}\tau'_{s}$ for the three
different values of $\eta$, which are very close to each other
according to Eq.~(\ref{critical2}).\label{fig2}}
\end{figure}

The phase velocity, defined as
$v_{\mathrm{p}}=\omega/|k_{\mathrm{r}}|$, of the spin accumulation
wave can be written as
\begin{equation}\label{vp}
\frac{v_{\mathrm{p}}}{c}=\frac{\sqrt{2}}{\eta+1}\left[
\sqrt{(\tilde{\omega}^{2}-\eta)^{2}+(\eta+1)^{2}\tilde{\omega}^{2}}
-(\tilde{\omega}^{2}-\eta)\right]^{\frac{1}{2}}.
\end{equation}
When $\tilde{\omega}\rightarrow0$ or $\infty$, the phase velocity
$v_{\mathrm{p}}$ will approach $2c/(\eta^{1/2}+\eta^{-1/2})$ or $c$,
respectively. When $\eta=1$, the phase velocity becomes equal to $c$
for all frequencies. Furthermore, the group velocity can be defined
as $v_{\mathrm{g}}=d\omega/dk_{\mathrm{r}}$ and calculated from
Eq.~(\ref{dispersion}).

Figure~\ref{fig2} shows the phase velocity $v_{\mathrm{p}}$ as
functions of $\tilde{\omega}$ for $\eta=0.1$, $0.02$, and $0.01$.
The phase velocity is approximately equal to the wavefront velocity
$c$ when the wave character is significant
($\omega>\omega_{\mathrm{c}}$). In this case, the phase velocity
provides a good description of the wave-like dynamics. On the
contrary, when the wave character is insignificant
($\omega<\omega_{\mathrm{c}}$), the wave amplitude is damped
strongly and the phase velocity is not meaningful any more. In this
region, the propagation velocity is the wavefront velocity $c$,
albeit only on the length scale of a damping length.

In the special case where $\eta=1$ ($\tau'_{s}=T_{1}$), we have
$|k_{\mathrm{r}}|=w/c$ and $|k_{\mathrm{i}}|=1/l_{\mathrm{sf}}$.
This means that the spin accumulation wave becomes a non-dispersive
but dissipative wave with the constant phase (and group) velocity
$c$ and penetration depth $l_{\mathrm{sf}}$. However, this case is
likely not realized because $T_{1}$ is usually much larger than
$\tau'_{s}$ and Valet-Fert theory is justified to be valid only when
$(\tau'_{s}/2{T}_{1})^{1/2}\ll1$.

\section{\label{sec5}Numerical results}

\begin{table}
\caption{\label{tab:table1}Parameters for Cu and Co used in
numerical calculation. The units of $v_{\mathrm{F}}$,
$\rho_{\mathrm{N(F)}}^{\ast}$, and $l_{\mathrm{sf}}^{\mathrm{N(F)}}$
are nm/ps, $\Omega\cdot$nm and nm, respectively. $\tau$ and $T_{1}$
are given in ps.}
\begin{ruledtabular}
\begin{tabular}{ccccccc}
% &\multicolumn{2}{c}{$D_{4h}^1$}&\multicolumn{2}{c}{$D_{4h}^5$}\\
 Material & $v_{\mathrm{F}}$ &
 $\rho_{\mathrm{N(F)}}^{\ast}$ &
 $l_{\mathrm{sf}}^{\mathrm{N(F)}}$ &
 $\tilde{\beta}$ & $\tau$ & $T_{1}$ \\ \hline
 Cu & $1570$\footnotemark[1] & $6$\footnotemark[2]  &
 $450$\footnotemark[2] & $0$ & $0.07$\footnotemark[5] & $3.5$\footnotemark[5] \\
 Co & $1570$\footnotemark[1] & $86$\footnotemark[3] &
 $60$\footnotemark[4]  & $0.5$\footnotemark[3] & $0.005$\footnotemark[5] & $0.9$\footnotemark[5] \\
\end{tabular}
\end{ruledtabular}
\footnotetext[1]{From Ref.~\onlinecite{Ash76}; Cu and Co are assumed
to have a common Fermi velocity in the Valet-Fert model.}
\footnotetext[2]{From Ref.~\onlinecite{Yang94}}
\footnotetext[3]{From Ref.~\onlinecite{vf}} \footnotetext[4]{From
Ref. \onlinecite{Piraux98}} \footnotetext[5]{Calculated from
Eqs.~(\ref{t1}) and (\ref{tau}).}
\end{table}

\begin{figure}
\includegraphics[width=.45\textwidth]{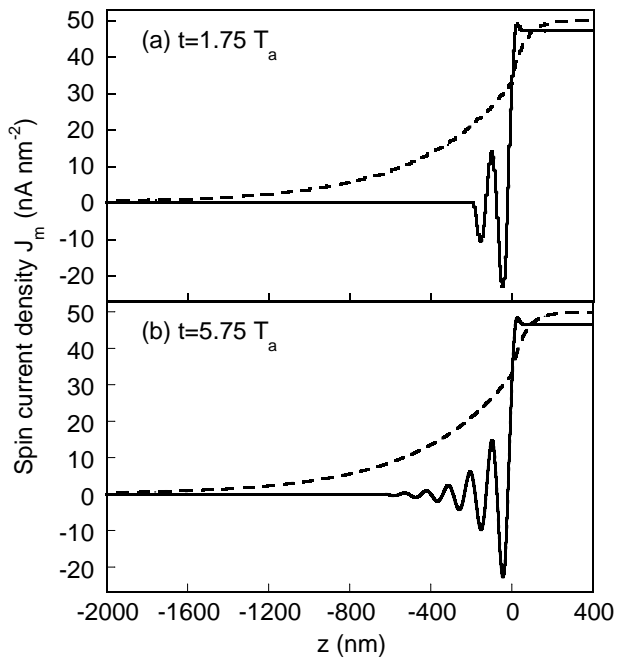}
\caption{Spin current density $J_{\mathrm{m}}(z,t)$ as a function of
  $z$. The solid curves in (a) and (b) are $J_{\mathrm{m}}(z,t)$ at
  $t=1.75~T_{\mathrm{a}}$ and $t=5.75~T_{\mathrm{a}}$ (charge current
  $J(t)=-J_{0}$) with AC drive, respectively. The dashed curves in (a)
  and (b) are the spin current density $J_{\mathrm{m}}(z)$ resulting
  from the DC current density, $J=-J_{0}$.
\label{fig3}}
\end{figure}

\begin{figure}[tb!]
\includegraphics[width=.45\textwidth]{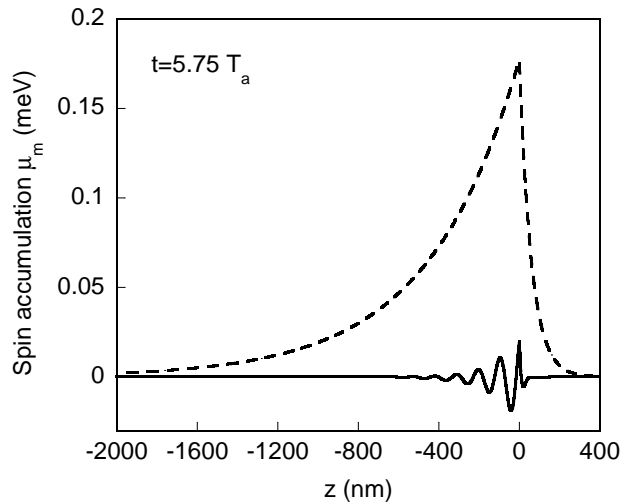}
\caption{Spin accumulation $\mu_{\mathrm{m}}(z,t)$ for the same
parameters as in Fig.~\ref{fig3}(b). \label{fig4}}
\end{figure}

In this section, the general analysis of the telegraph equations for
spin-transport is augmented by numerical solutions for two
illustrative examples of signal propagation using spin polarized
currents through a ferromagnet/metal junction: (i) injection of an
alternating current, and (ii) instantaneous magnetization switching.
The results are obtained by numerically solving the system of
Eqs.~(\ref{new3}) and (\ref{new4}). Our numerical method is outlined
in Appendix~\ref{numerical}. These equations are equivalent to the
telegraph equations Eqs.~(\ref{new5}) and (\ref{new6}), which have
been discussed in the previous section, but are easier to solve.
Alternatively, we could solve the equation system consisting of
Eqs.~(\ref{new7}) and (\ref{new8}), in which the spin accumulation
is described by the spin density. However, it is more convenient to
work with the electrochemical potential than the spin density when
we deal with the boundary conditions.\cite{Her97}

We choose a ferromagnet/metal junction consisting of Co and Cu as the
material system in both of the scenarios. The interface of the
junction is placed at $z=0$ and the Co (Cu) occupies the half-space
$z>0$ ($z<0$). The positive direction of the current is parallel to
the positive direction of the $z$ axis. For simplicity, the interface
resistance of the junction will be neglected. Then, the
electrochemical potential and the current density are continuous
across the interface. Consequently, the spin transport across two
layers can be described by one common equation system with different
material parameters for the two layers.

The material parameters used in our numerical calculation are shown in
Table~\ref{tab:table1}. All other parameters can be obtained from the
values in Tab.~\ref{tab:table1}. In particular, the wavefront velocity
is calculated to be $c=910$\,nm/ps from Eq.~(\ref{wfvelo}). In the
nonmagnetic layer, $\eta=\tau'_{s}/T_{1}=0.02$. The wavelength
$\lambda$ and damping length $l_{\mathrm{d}}$ are shown as the solid
curves in Fig.~\ref{fig1}. The critical period $T_{\mathrm{c}}$ can be
estimated to be $2\tau'_{s}\simeq 0.14$\,ps from
Eq.~(\ref{critical2}), and the phase velocity is plotted in
Fig.~\ref{fig2}.

\subsection{AC current injection}

The alternating charge current density passing through the
ferromagnet/metal junction is assumed to be of the form
$J(t)=J_{0}\sin(\omega t)$, where $J_{0}=100$ nA/nm$^{2}$. Note that
the $z$-dependence of the charge current $J(z,t)$ in Eq.~(\ref{new4})
is neglected for the investigation of the spin transport as pointed
out in Sec.~\ref{sec2}. Two typical frequencies are studied in the
case of the AC drive: $\nu_{\mathrm{a}}=\omega_{a}/(2\pi)=8.33$\,THz
and $\nu_{\mathrm{b}}=\omega_{b}/(2\pi)=0.23$\,THz, which are larger
and smaller than the critical frequency 
$\nu_{\mathrm{c}}=\omega_{\mathrm{c}}/(2\pi)=7.11$\,THz of Cu,
respectively. The corresponding periods of the two frequencies are
$T_{\mathrm{a}}=0.12$\,ps and $T_{\mathrm{b}}=4.4$\,ps, which satisfy
$T_{\mathrm{a}}<T_{\mathrm{c}}<T_{\mathrm{b}}$. The numerical results
for the two frequencies are discussed in the following.

\begin{figure}
\includegraphics[width=.45\textwidth]{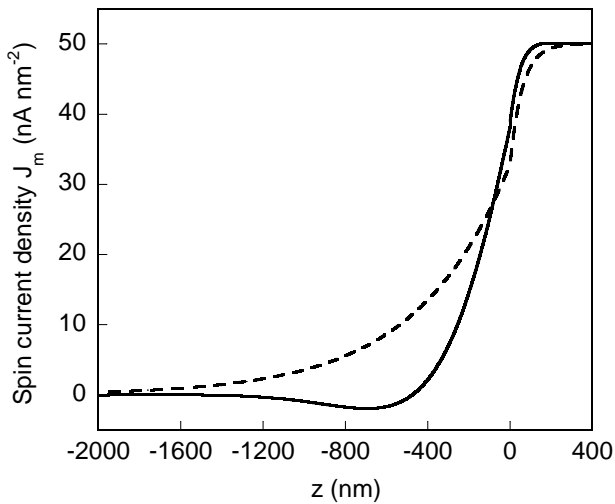}
\caption{Spin current density $J_{\mathrm{m}}(z,t)$ as a function of
  $z$. The solid curve is $J_{\mathrm{m}}(z,t)$ at
  $t=1.75~T_{\mathrm{b}}$ (charge current $J(t)=-J_{0}$) with AC
  drive. The dashed curve is the spin current density
  $J_{\mathrm{m}}(z)$ for the case of a DC current density
  $J=-J_{0}$. \label{fig5}}
\end{figure}

\begin{figure}
\includegraphics[width=.45\textwidth]{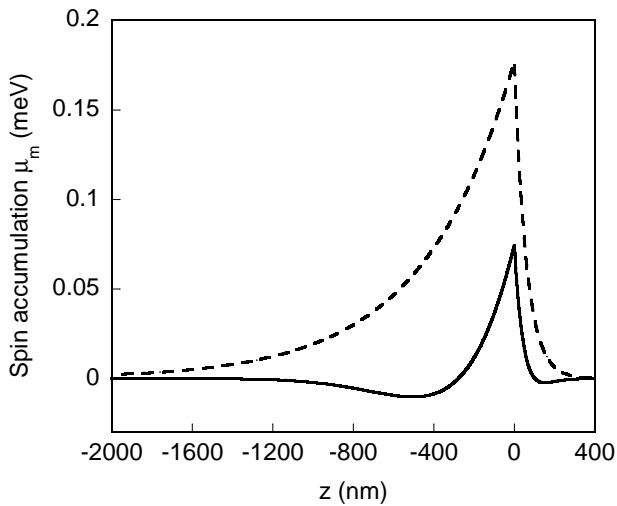}
\caption{Spin accumulation $\mu_{\mathrm{m}}(z,t)$ for the same
parameters as in Fig.~\ref{fig5}. \label{fig6}}
\end{figure}

\emph{High-frequency case ($\omega = \omega_{\mathrm{a}}>
  \omega_{\mathrm{c}}$):}\, Figure~\ref{fig3} shows snapshots of the
spin current density $J_{\mathrm{m}}(z,t)$ at times
$t=1.75\ T_{\mathrm{a}}$ and $t=5.75\ T_{\mathrm{a}}$. At both times,
the charge current density $J(t)$ reaches its minimum
$J(t)=-J_{0}$. The wavefront, i.e., the spin signal, can be seen
clearly in Fig.~\ref{fig3}(a), where the time $t$ is so small that the
wavefront has not propagated beyond the scale of the damping length
$l_{\mathrm{d}}$. In Fig.~\ref{fig3}(b), the signal has propagated
further, and due to the attenuation of the wave, the wave front is
less clearly visible. Nevertheless, the wavefront velocity $c$ can be
determined numerically (or experimentally) by tracking the motion of
the wavefront over a short time interval after switching on the drive
current. Since we are using a signal time scale shorter than the
critical time, we expect from the analysis in Sec.~\ref{sec3} (see
also Fig.~\ref{fig2}) the phase velocity to be $v_{\mathrm{p}}\simeq
c=$910\,nm/ps from Eq.~(\ref{wfvelo}), and a wavelength
$\lambda=108$\,nm. These expectations are borne out by the numerical
results. The dynamical damping length $l_{\mathrm{d}}$ can also be
extracted from the numerical data, or from an experiment, by fitting a
decay time to the envelope of the spin-current signal for longer
times. Due to inaccuracies of the fitting procedure, this quantity is
more difficult to determine quantitatively, but agrees well with the
damping length $l_{\mathrm{d}}=126$\,nm expected from Eq~(\ref{dl}). An
important qualitative conclusion can be drawn by comparing the decay
of the dynamical spin signal in Fig.~\ref{fig3}(b) with the spin
current density $J_{\mathrm{m}}(z)$ that results from a DC current
density $J=-J_{0}$, which is also shown. Since our dynamical equations
and the spin-diffusion equation have the same long-time limit, the DC
result is identical with steady-state spin diffusion. It is apparent
that the damping length $l_{\mathrm{d}}$ becomes much shorter than the
spin diffusion length $l_{\mathrm{sf}}$ of the steady-state spin
transport with DC bias. This is the ``skin'' effect, which is already
present in the analytical results in Sec.~\ref{sec3}.

Figure~\ref{fig4} shows the $z$-dependent spin accumulation
$\mu_{\mathrm{m}}$ for the same parameters as in Fig.~\ref{fig3}(b).
The wavelength, damping length, and phase velocity given by
Fig.~\ref{fig4} are very similar to those in Fig.~\ref{fig3}(b).
Note, however, that the amplitude of the dynamical spin accumulation
is much smaller than the spin accumulation of the steady-state spin
transport shown by the dashed curve. The reason is that the AC drive
oscillates too fast so that the spin accumulation does not have enough
time to reach its steady-state value.

\begin{figure}
\includegraphics[width=.45\textwidth]{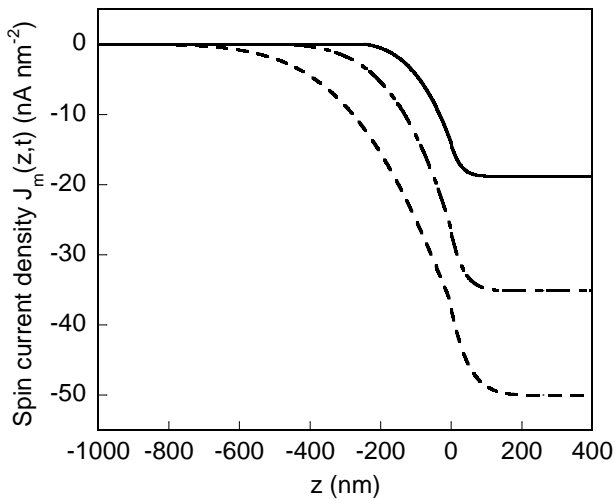}
\caption{Spin current density $J_{\mathrm{m}}(z,t)$ as a function of
$z$. The solid, dot-dashed, and dashed curves are
$J_{\mathrm{m}}(z,t)$ at $t=T_{\mathrm{b}}/16,\ T_{\mathrm{b}}/8,\
T_{\mathrm{b}}/4$, respectively, with AC drive. \label{fig7}}
\end{figure}

\begin{figure}
\includegraphics[width=.45\textwidth]{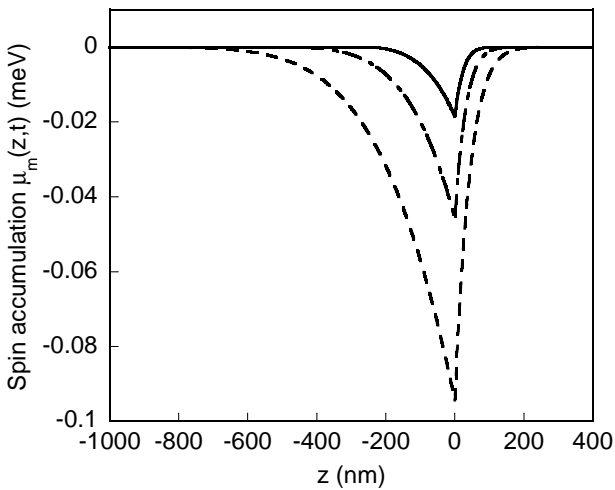}
\caption{Spin accumulation $\mu_{\mathrm{m}}(z,t)$ for the same
parameters as in Fig.~\ref{fig7}. \label{fig8}}
\end{figure}

\emph{Low-frequency case ($\omega = \omega_{\mathrm{b}} <
  \omega_{\mathrm{c}}$):} Before analyzing the signal propagation
velocity, we first show how the results change qualitatively
compared with the high-frequency case.  In Figure~\ref{fig5}, the
spin current $J_{\mathrm{m}}$ is plotted as a function of $z$ driven
by an AC current with frequency $\omega_{\mathrm{b}}$, which is
smaller than the critical angular frequency $\omega_{\mathrm{c}}$. The
period $T_{\mathrm{b}}$ of the AC drive is $4.4$\,ps, which is much
larger than $T_{\mathrm{a}}$ in Fig.~\ref{fig3}. The solid curve is
$J_{\mathrm{m}}(z,t)$ at $t=1.75 T_{\mathrm{b}}$ (charge current
$J=-J_{0}$) with AC drive.  The dashed curve is again
$J_{\mathrm{m}}(z)$ driven by a DC current density $J=-J_{0}$. For
this driving frequency, the wave character is insignificant, because
the wavelength $\lambda=1856$ nm becomes much larger than the
damping length $l_{\mathrm{d}}=268$ nm, and thus the wave amplitude
is damped to zero within just one wavelength. From a practical point
of view, the wavelength and the phase velocity $v_{\mathrm{p}}=0.47
c$ lose their meaning in this case.  Comparison between the solid
and dashed curves shows that the damping length for $T_{\mathrm{b}}$
becomes longer than that for $T_{\mathrm{a}}$ in Fig.~\ref{fig3},
which is a consequence of the ``skin'' effect.

Figure~\ref{fig6} shows the spin accumulation $\mu_{\mathrm{m}}$ as
a function of $z$. The parameters used are the same as those in
Fig.~\ref{fig5}. The features of the spin accumulation are
again reminiscent of the spin current in Fig.~\ref{fig5}. Note
that the spin accumulation has become larger compared with the AC
drive with period $T_{\mathrm{a}}$ in Fig.~\ref{fig4}. This is
reasonable because the AC drive oscillates more slowly than that in
Fig.~\ref{fig4}, so that the spin accumulation has more time to
approach its steady-state value.

At the time the snapshots in Figs.~\ref{fig5} and~\ref{fig6} are taken,
no ``wave front'' of the spin current, or signal, can be
distinguished. To determine the propagation velocity, we show
$J_{\mathrm{m}}(z,t)$ and $\mu_{\mathrm{m}}(z,t)$ at
$t=T_{\mathrm{b}}/16$, $T_{\mathrm{b}}/8$, and $T_{\mathrm{b}}/4$ in
Figs.~\ref{fig7} and \ref{fig8}, respectively. By tracking the motion
of the wavefront with time, we can estimate the propagation velocity
of the signal. The result is in agreement with the ``wave front''
velocity $c$, which according to our analysis of the telegraph
equation~(\ref{new5}) is still the propagation velocity.

Time-dependent spin transport in the low-frequency case can be
described approximately by the conventional spin diffusion
equation. However, it is impossible to estimate the signal propagation
velocity from conventional spin diffusion theory because there
is no wavefront in that case and the signal appears in infinity once
the charge current $J(t)$ is switched on.~\cite{zhang02,Mas96}

\begin{figure}{tb}
\includegraphics[width=.45\textwidth]{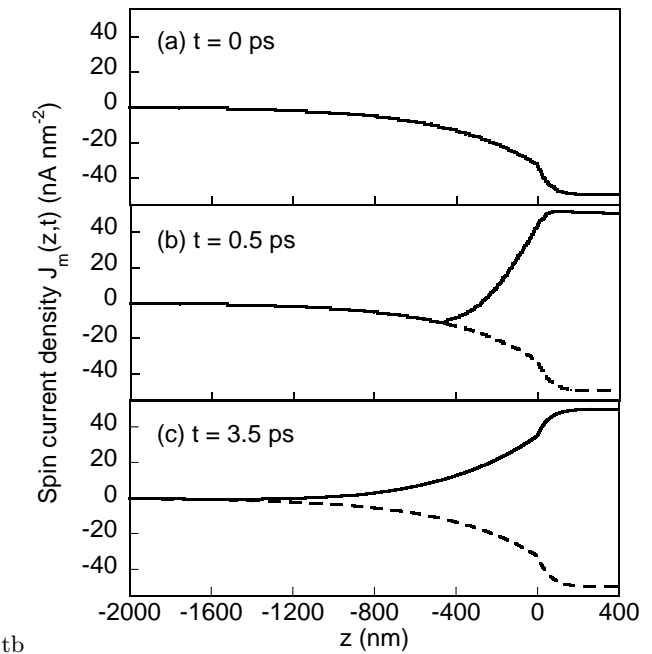}
\caption{Spin current density $J_{\mathrm{m}}(z,t)$ as a function of
$z$. The solid curves in Figs.~(a), (b) and (c) are
$J_{\mathrm{m}}(z,t)$ at $t=0,\ 0.5,\ 3.5$~ps, respectively. The
dashed curves in (b) and (c) are $J_{\mathrm{m}}(z,t)$ at $t=0$~ps
plotted again as a reference. \label{fig9}}
\end{figure}

\begin{figure}{tb}
\includegraphics[width=.45\textwidth]{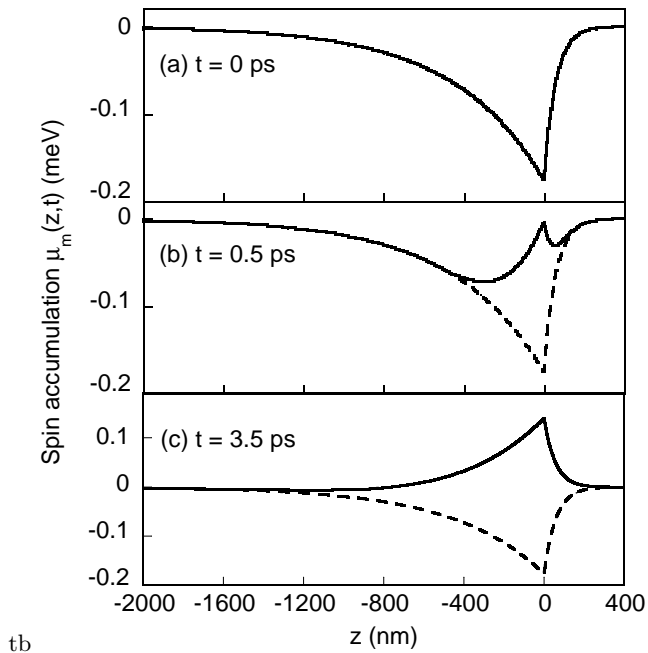}
\caption{Spin accumulation $\mu_{\mathrm{m}}(z,t)$ for the same
parameters as in Fig.~\ref{fig9}. \label{fig10}}
\end{figure}

\subsection{Magnetization switching}

The instantaneous switching of the magnetization in the ferromagnet,
through which the current passes into the nonmagnetic metal, provides
perhaps the conceptually cleanest picture of a spin-switching process.
For a numerical study of this process, we consider again a
ferromagnet/metal junction consisting of Co and Cu. We assume that the
system is in a steady state in the presence of the DC drive with a
charge current density $J_{0}=100$~nA/nm$^{2}$ before the
magnetization of the ferromagnetic layer is switched from ``up'' to
``down'' at $t=0$. We model the switching as an idealized
instantaneous process, and only consider the evolution of the spin
current density and spin accumulation afterwards. The spin-up
electrons become the majority, and the spin-down electrons become the
minority after the instantaneous switching. The conductivities of the
majority and minority channels are also exchanged by the switching.
Although the evolution of $J_{\mathrm{m}}(z,t)$ and
$\mu_{\mathrm{m}}(z,t)$ does not take the waveform used in
Sec.~\ref{sec3}, it can be decomposed into different frequencies by
Fourier transformation, so that the analysis of the telegraph equation
still applies.

Figures~\ref{fig9} and \ref{fig10} show the dynamics of the spin
current and spin accumulation. Starting from the steady-state value
shown in parts (a), the magnetization is switched instantaneously at
$t=0$. Figs.~\ref{fig9}(b) and Fig.~\ref{fig10}(b) show snapshots
0.5\,ps after the switch, when a pronounced kink has developed. This
kink indicates the leftmost position to which the
magnetization-switching signal has propagated after 0.5\,ps. The
kink is noticeable only if the time $t$ is so small that it does not
propagate beyond the length scale of the spin-diffusion length
$l_{\mathrm{sf}}$, over which the steady-state signal decays. Thus
the signal-propagation velocity can be estimated roughly by tracking
the motion of the kink with time at the early stage of the
switching. The result is very close to the wavefront velocity
$c=910$~nm/ps calculated from the analytical result,
Eq.~(\ref{wfvelo}). Moreover, Fig.~\ref{fig9}(c) shows that the spin
current density reaches the steady state with ``down'' magnetization
on the time scale of the spin relaxation time $T_{1}$.  Since
$T_{1}\gg\tau'_{s}$, we can consider $t=T_{1}$ as the long-time
limit. This behavior is consistent with the result calculated from
the diffusion equation in Ref.~\onlinecite{zhang02}, so that again
the diffusion character of spin transport emerges as an
approximation of the wave-diffusion character in the long-time
limit.

\section{\label{sec6}Summary}

We studied signal propagation in time-dependent spin transport through
magnetic multilayers using an extension of the Valet-Fert theory to
time-dependent phenomena. We established that time-dependent spin
transport has a wave character in addition to its diffusive character,
which enabled us to determine the finite propagation velocity of
signals in spin transport, such as AC spin injection and magnetization
switching. The propagation velocity is the wavefront velocity
$c=v_{\mathrm{F}}/\sqrt{3}$. The wave character is significant if the
signal time scale $\tau_{\mathrm{sig}}$ is smaller than a critical
time $T_{\mathrm{c}}$. When the wave character is significant
($\tau_{\mathrm{sig}}<T_{\mathrm{c}}$), the time-dependent spin
transport should be modeled by the dynamical equations introduced in
this paper, or, equivalently, the telegraph equations. However, pure
diffusive spin transport can be regarded as an approximation of the
wave-diffusion duality for slow switching times
($\tau_{\mathrm{sig}}\gg{T}_{\mathrm{c}}$). In this limit, the spin
diffusion equation can be used to study the time-dependence of spin
transport approximately, but it incorrectly yields an infinite
signal-propagation velocity.

\begin{acknowledgments}
We acknowledge financial support from the state of
Rheinland-Pfalz through the MATCOR program, and a
CPU-time grant from the John von Neumann Institut for
Computing (NIC) at the Forschungszentrum J\"{u}lich.
\end{acknowledgments}

\appendix

\section{\label{appendix}Identities and derivations}

\subsection{Useful identities}

Several useful identities will be established by the help of
Eq.~(\ref{lp}). Multiplying $\sin\theta$ and integrating over
$\theta$ from $0$ to $\pi$ on both sides of Eq.~(\ref{lp}), we have
\begin{equation}\label{id1}
\int_{0}^{\pi}d\theta\sin\theta{g}_{s}(z,\mathbf{v},t)
\!=\!\sum_{n=1}^{\infty}g_{s}^{(n)}(z,t)\int_{0}^{\pi}d\theta\sin\theta
{P}_{n}(\cos\theta).
\end{equation}
The right-hand-side (RHS) of Eq.~(\ref{id1}) can be further written as
\begin{equation}\label{id2}
RHS=\sum_{n=1}^{\infty}g_{s}^{(n)}(z,t)\int_{-1}^{1}d{u}P_{0}(u)
P_{n}(u).
\end{equation}
Using the orthogonality relation between Legendre polynomials
\begin{equation}
\int_{-1}^{1}duP_{n'}(u)P_{n}(u)=\frac{2}{2n+1}\delta_{n,n'},
\end{equation}
where $\delta_{n,n'}$ is the usual Kronecker symbol, we
obtain from Eqs.~(\ref{id1}) and (\ref{id2})
\begin{equation}\label{id3}
\int_{0}^{\pi}d\theta\sin\theta{g}_{s}(z,\mathbf{v},t)=0.
\end{equation}
Eq.~(\ref{id3}) further leads to
\begin{equation}\label{id4}
\begin{split}
&\sum_{\mathbf{v}}g_{s}(z,\mathbf{v},t)=\frac{Vm^{3}}{h^{3}}\int{d}^{3}vg_{s}(z,\mathbf{v},t)\\
&=\frac{Vm^{3}}{h^{3}}\int_{0}^{2\pi}
\!\!\!d\varphi\int_{0}^{\pi}\!\!d\theta\sin\theta\!\int_{0}^{\infty}\!\!\!dvv^{2}g_{s}(z,\mathbf{v},t)
=0.
\end{split}
\end{equation}

\subsection{\label{derrta}Derivation of Eq.~(\ref{rta})}

Following Ref.~\onlinecite{vf}, we substitute Eq.~(\ref{appr}) into Eq.~(\ref{bte}) and use the
following identity
\begin{equation}
\frac{\partial{f}^{0}}{\partial\varepsilon}=\frac{1}{mv}\frac{\partial{f}^{0}}{\partial{v}}
=\frac{-\delta(v-v_{\mathrm{F}})}{mv_{\mathrm{F}}}.
\end{equation}
Then we can write the RHS (the collision terms) of Eq.~(\ref{appr}) as
\begin{widetext}
\begin{equation}\label{scatt1}
\begin{split}
\frac{\partial{f}_{s}(z,\mathbf{v},t)}{\partial
t}\bigg|_{\rm{collision}}
=-\frac{\partial{f}^{0}(v)}{\partial\varepsilon}&
P_{s}[z,\varepsilon(v)]\frac{4\pi{v}}{m}\bigg[g_{s}(z,\mathbf{v},t)
-\frac{1}{2}\int_{0}^{\pi}d\theta'\sin\theta'{g}_{s}(z,\mathbf{v}',t)\big|_{v'=v}\bigg]\\
-\frac{\partial{f}^{0}(v)}{\partial\varepsilon}&
P_{\rm{sf}}[z,\varepsilon(v)]\frac{4\pi{v}}{m}\bigg[g_{s}(z,\mathbf{v},t)
-\frac{1}{2}\int_{0}^{\pi}d\theta'\sin\theta'{g}_{-s}(z,\mathbf{v}',t)\big|_{v'=v}\bigg]\\
+\frac{\partial{f}^{0}(v)}{\partial\varepsilon}&P_{\rm{sf}}[z,\varepsilon(v)]\frac{4\pi{v}}{m}
[\mu_{s}(z,t)-\mu_{-s}(z,t)].
\end{split}
\end{equation}

Using Eq.~(\ref{id3}), we can write Eq.~(\ref{scatt1}) in the form
\begin{equation}\label{scatt2}
\begin{split}
\frac{\partial{f}_{s}(z,\mathbf{v},t)}{\partial
t}\bigg|_{\rm{collision}}
=-\frac{\partial{f}^{0}(v)}{\partial\varepsilon}&
P_{s}[z,\varepsilon(v)]\frac{4\pi{v}}{m}g_{s}(z,\mathbf{v},t)
-\frac{\partial{f}^{0}(v)}{\partial\varepsilon}
P_{\rm{sf}}[z,\varepsilon(v)]\frac{4\pi{v}}{m}g_{s}(z,\mathbf{v},t)\\
+\frac{\partial{f}^{0}(v)}{\partial\varepsilon}&P_{\rm{sf}}
[z,\varepsilon(v)]\frac{4\pi{v}}{m}[\mu_{s}(z,t)-\mu_{-s}(z,t)].
\end{split}
\end{equation}
\end{widetext}
By introducing the relaxation times
\begin{eqnarray}
\frac{1}{\tau_{s}(v)}&=&P_{s}[z,\varepsilon(v)]\frac{4\pi{v}}{m},\label{rtas}\\
\frac{1}{\tau_{\rm{sf}}(v)}&=&P_{\rm{sf}}[z,\varepsilon(v)]\frac{4\pi{v}}{m},\label{rtasf}
\end{eqnarray}
where the $z$-dependence of the relaxation times is neglected within
the same layer, we can further write Eq.~(\ref{scatt2}) as
\begin{equation}
\begin{split}
\frac{\partial f_{s}(z,\mathbf{v},t)}{\partial
t}&\bigg|_{\rm{collision}}\\
=-&\frac{\partial{f}^{0}(v)}{\partial\varepsilon}
\left(\frac{1}{\tau_{s}}+\frac{1}{\tau_{\rm{sf}}}\right)g_{s}(z,\mathbf{v},t)\\
+&\frac{\partial{f}^{0}(v)}{\partial\varepsilon}
\frac{\mu_{s}(z,t)-\mu_{-s}(z,t)}{\tau_{\rm{sf}}}.
\end{split}
\end{equation}
Taking into account the left-hand-side of Eq.~(\ref{bte}) and
integrating over $v$, we can finally derive Eq.~(\ref{rta}). Note
that $\tau_{s}(v)$ and $\tau_{\mathrm{sf}}(v)$ are restricted to the
Fermi velocity $v_{\mathrm{F}}$ after the integration over $v$ and
then they are simply written as $\tau_{s}$ and $\tau_{\mathrm{sf}}$.

\subsection{\label{derden}Derivation of Eqs.~(\ref{den3}) and (\ref{total})}

Multiplying both sides of Eq.~(\ref{appr}) by $-e/V$, summing over
$\mathbf{v}$, and using Eq.~(\ref{id4}), we obtain
\begin{equation}\label{den}
n_{s}(z,t)-n_{s}^{0}=-eN_{s}[\mu_{s}(z,t)-\mu^{0}],
\end{equation}
where
\begin{eqnarray}
n_{s}(z,t)&=&-\frac{e}{V}\sum_{\mathbf{v}}f_{s}(z,\mathbf{v},t),\\
n_{s}^{0}&=&-\frac{e}{V}\sum_{\mathbf{v}}f^{0}(v)=-en_{s},\label{density}
\end{eqnarray}
are the nonequilibrium and equilibrium charge density for spin $s$,
respectively. In turn, Eq.~(\ref{den}) yields
\begin{eqnarray}
n_{\mathrm{m}}(z,t)&=&-eN_{s}\mu_{\mathrm{m}}(z,t),\quad\label{nmu}\\
n(z,t)-2n_{s}^{0}&=&-eN_{s}\left[2\mu(z,t)-2\mu^{0}\right],\quad\label{nmu2}
\end{eqnarray}
where $n_{\mathrm{m}}(z,t)=n_{+}(z,t)-n_{-}(z,t)$ is the spin
density and $n(z,t)=n_{+}(z,t)+n_{-}(z,t)$ the total nonequilibrium
charge density.

\subsection{\label{numerical}Numerical solution of Eqs.~(\ref{new3}) and (\ref{new4})}

For the numerical solution of Eqs.~(\ref{new3}) and (\ref{new4}) we
use the method of characteristics and Hartree's computational form.
Following Ref.~\onlinecite{Ames92}, the space $z$ and time $t$ are
discretized into grids with equal intervals $\Delta{z}$ and
$\Delta{t}$, respectively. The discretized forms of
$J_{\mathrm{m}}(z,t)$ and $\mu_{\mathrm{m}}(z,t)$ are
$J_{\mathrm{m},i}^{n}$ and $\mu_{\mathrm{m},i}^{n}$ at $i$th space
point and $n$th time point, respectively. Then,
$J_{\mathrm{m},i}^{n+1}$ and $\mu_{\mathrm{m},i}^{n+1}$ at $(n+1)$th
time point can be calculated by the iteration relations
\begin{equation}\label{new3d}
\begin{split}
\bigg(2+\frac{\Delta{t}}{T_{1}}\bigg)\mu_{\textrm{m},i}^{n+1}
=\bigg(1-\frac{\Delta{t}}{2T_{1}}\bigg)(\mu_{\textrm{m},i-1}^{n}
+\mu_{\textrm{m},i+1}^{n})\\
-\frac{1}{eN_{s}c}\bigg(1-\frac{\Delta t}{2\tau}\bigg)
(J_{\textrm{m},i-1}^{n}-J_{\textrm{m},i+1}^{n}),
\end{split}
\end{equation}

\begin{equation}\label{new4d}
\begin{split}
\bigg(2+\frac{\Delta{t}}{\tau}\bigg)J_{\textrm{m},i}^{n+1}
=-eN_{s}c\bigg(1-\frac{\Delta{t}}{2T_{1}}\bigg)
(\mu_{\textrm{m},i-1}^{n}-\mu_{\textrm{m},i+1}^{n})\\
+\bigg(1-\frac{\Delta{t}}{2\tau}\bigg)
(J_{\textrm{m},i-1}^{n}+J_{\textrm{m},i+1}^{n})
-\frac{\Delta{t}}{\tau}\tilde{\beta}(J^{n}+J^{n+1}),
\end{split}
\end{equation}
for all space points except the two boundary points, which should be
determined by boundary conditions. Here, $J^{n}$ is the total current
density at $n$th time point. Moreover, $\Delta{z}$ and $\Delta{t}$ are
chosen to satisfy the relation
$\Delta{z}=c\Delta{t}$. Eqs.~(\ref{new3d}) and (\ref{new4d}) can be
iterated numerically to yield the results presented in
Sec.~\ref{sec5}. In the numerical solution, we used the following
initial and boundary conditions for the AC spin injection and
magnetization switching.

\emph{AC spin injection}: The initial conditions are 
$\mu_{\mathrm{m}}(z,t=0)=0$ and $J_{\mathrm{m}}(z,t=0)=0$. The
boundary condition for $\mu_{\mathrm{m}}(z,t)$ is
$\mu_{\mathrm{m}}(z=\pm\infty,t)=0$. From Eq.~(\ref{new4}), the boundary condition
 
\begin{equation}
\begin{split}
J_{\mathrm{m}}(z=\pm\infty,t)=\frac{\tilde{\beta}J_{0}}{1+\omega^{2}\tau^{2}}
[\omega\tau\cos(\omega{t})-\sin(\omega{t})\\
-\omega\tau\exp(-t/\tau)].
\end{split}
\end{equation}
for $J_{\mathrm{m}}(z,t)$ can be derived.

\emph{Magnetization switching}: The initial conditions for
$J_{\mathrm{m}}(z,t)$ and $\mu_{\mathrm{m}}(z,t)$ are the
steady-state solutions to Eqs.~(\ref{new3}) and (\ref{new4})
\begin{eqnarray}
\mu_{\mathrm{m}}^{\mathrm{F}}(z,t=0)&=&C_{0}\exp(-z/l_{\mathrm{sf}}^{\mathrm{F}}),\\
J_{\mathrm{m}}^{\mathrm{F}}(z,t=0)&=&-\frac{C_{0}}
{2e\rho_{\mathrm{F}}^{\ast}l_{\mathrm{sf}}^{\mathrm{F}}}
\exp(-z/l_{\mathrm{sf}}^{\mathrm{F}})-\tilde{\beta}J_{0},\\
\mu_{\mathrm{m}}^{\mathrm{N}}(z,t=0)&=&C_{0}\exp(z/l_{\mathrm{sf}}^{\mathrm{N}}),\\
J_{\mathrm{m}}^{\mathrm{N}}(z,t=0)&=&\frac{C_{0}}{2e\rho_{\mathrm{N}}^{\ast}
l_{\mathrm{sf}}^{\mathrm{N}}}\exp(z/l_{\mathrm{sf}}^{\mathrm{N}}),
\end{eqnarray}
where
$C_{0}=-2e\tilde{\beta}J_{0}(\rho_{\mathrm{F}}^{\ast}l_{\mathrm{sf}}^{\mathrm{F}}
\rho_{\mathrm{N}}^{\ast}l_{\mathrm{sf}}^{\mathrm{N}})
/(\rho_{\mathrm{F}}^{\ast}l_{\mathrm{sf}}^{\mathrm{F}}+\rho_{\mathrm{N}}^{\ast}
l_{\mathrm{sf}}^{\mathrm{N}})$. Here, $\mu_{\mathrm{m}}^{\mathrm{F}}$
and $J_{\mathrm{m}}^{\mathrm{F}}$ apply to the ferromagnetic layer
occupying $z>0$, whereas $\mu_{\mathrm{m}}^{\mathrm{N}}$ and
$J_{\mathrm{m}}^{\mathrm{N}}$ refer to the nonmagnetic layer ($z<0$).
In deriving the initial conditions above, we have used the identity
$1/(2\rho_{\mathrm{N(F)}}^{\ast})=\sigma_{\mathrm{N(F)}}/2
=e^{2}N_{s}\bar{D}_{\mathrm{N(F)}}$, where $\sigma_{\mathrm{N(F)}}$ is
the total conductivity of the nonmagnetic (ferromagnetic) layer.  The
boundary condition for $\mu_{\mathrm{m}}(z,t)$ is
$\mu_{\mathrm{m}}(z=\pm\infty,t)=0$. Then, the boundary condition for
$J_{\mathrm{m}}(z,t)$ can again be derived from Eq.~(\ref{new4}). This
yields
\begin{equation}
J_{\mathrm{m}}(z=\pm\infty,t)=\tilde{\beta}J_{0}[1-2\exp(-t/\tau)],
\end{equation}
where $\tilde{\beta}$ is the asymmetry parameter before the
magnetization switching. Note that $\tilde{\beta}$ becomes
$-\tilde{\beta}$ when the magnetization is switched ($t>0$).

\bibliography{references_corrected}% Produces the bibliography via BibTeX.

\begin{thebibliography}{21}
\expandafter\ifx\csname natexlab\endcsname\relax\def\natexlab#1{#1}\fi
\expandafter\ifx\csname bibnamefont\endcsname\relax
  \def\bibnamefont#1{#1}\fi
\expandafter\ifx\csname bibfnamefont\endcsname\relax
  \def\bibfnamefont#1{#1}\fi
\expandafter\ifx\csname citenamefont\endcsname\relax
  \def\citenamefont#1{#1}\fi
\expandafter\ifx\csname url\endcsname\relax
  \def\url#1{\texttt{#1}}\fi
\expandafter\ifx\csname urlprefix\endcsname\relax\def\urlprefix{URL }\fi
\providecommand{\bibinfo}[2]{#2}
\providecommand{\eprint}[2][]{\url{#2}}

\bibitem[{\citenamefont{\v{Z}uti\'{c} et~al.}(2004)\citenamefont{\v{Z}uti\'{c},
  Fabian, and Sarma}}]{zut04}
\bibinfo{author}{\bibfnamefont{I.}~\bibnamefont{\v{Z}uti\'{c}}},
  \bibinfo{author}{\bibfnamefont{J.}~\bibnamefont{Fabian}}, \bibnamefont{and}
  \bibinfo{author}{\bibfnamefont{S.~D.} \bibnamefont{Sarma}},
  \bibinfo{journal}{Rev. Mod. Phys.} \textbf{\bibinfo{volume}{76}},
  \bibinfo{pages}{323} (\bibinfo{year}{2004}).

\bibitem[{\citenamefont{Fabian et~al.}(2007)\citenamefont{Fabian,
  Matos-Abiague, Ertler, Stano, and \v{Z}uti\'{c}}}]{Fab07}
\bibinfo{author}{\bibfnamefont{J.}~\bibnamefont{Fabian}},
  \bibinfo{author}{\bibfnamefont{A.}~\bibnamefont{Matos-Abiague}},
  \bibinfo{author}{\bibfnamefont{C.}~\bibnamefont{Ertler}},
  \bibinfo{author}{\bibfnamefont{P.}~\bibnamefont{Stano}}, \bibnamefont{and}
  \bibinfo{author}{\bibfnamefont{I.}~\bibnamefont{\v{Z}uti\'{c}}},
  \bibinfo{journal}{Acta Phys. Slovaca} \textbf{\bibinfo{volume}{57}},
  \bibinfo{pages}{565} (\bibinfo{year}{2007}).

\bibitem[{\citenamefont{Zhang and Levy}(2002)}]{zhang02}
\bibinfo{author}{\bibfnamefont{S.}~\bibnamefont{Zhang}} \bibnamefont{and}
  \bibinfo{author}{\bibfnamefont{P.~M.} \bibnamefont{Levy}},
  \bibinfo{journal}{Phys. Rev. B} \textbf{\bibinfo{volume}{65}},
  \bibinfo{pages}{052409} (\bibinfo{year}{2002}).

\bibitem[{\citenamefont{Rashba}(2002)}]{Rash02}
\bibinfo{author}{\bibfnamefont{E.~I.} \bibnamefont{Rashba}},
  \bibinfo{journal}{Appl. Phys. Lett.} \textbf{\bibinfo{volume}{80}},
  \bibinfo{pages}{2329} (\bibinfo{year}{2002}).

\bibitem[{\citenamefont{Zhang and Levy}(2005)}]{zhang05}
\bibinfo{author}{\bibfnamefont{J.}~\bibnamefont{Zhang}} \bibnamefont{and}
  \bibinfo{author}{\bibfnamefont{P.~M.} \bibnamefont{Levy}},
  \bibinfo{journal}{Phys. Rev. B} \textbf{\bibinfo{volume}{71}},
  \bibinfo{pages}{184417} (\bibinfo{year}{2005}).

\bibitem[{\citenamefont{Cywi\'{n}ski et~al.}(2006)\citenamefont{Cywi\'{n}ski,
  Dery, and Sham}}]{cy06}
\bibinfo{author}{\bibfnamefont{{\L}.}~\bibnamefont{Cywi\'{n}ski}},
  \bibinfo{author}{\bibfnamefont{H.}~\bibnamefont{Dery}}, \bibnamefont{and}
  \bibinfo{author}{\bibfnamefont{L.~J.} \bibnamefont{Sham}},
  \bibinfo{journal}{Appl. Phys. Lett.} \textbf{\bibinfo{volume}{89}},
  \bibinfo{pages}{042105} (\bibinfo{year}{2006}).

\bibitem[{\citenamefont{Maxwell}(1867)}]{Max67}
\bibinfo{author}{\bibfnamefont{J.~C.} \bibnamefont{Maxwell}},
  \bibinfo{journal}{Phil. Trans. Roy. Soc.} \textbf{\bibinfo{volume}{157}},
  \bibinfo{pages}{49} (\bibinfo{year}{1867}).

\bibitem[{\citenamefont{Cattaneo}(1948)}]{Cat48}
\bibinfo{author}{\bibfnamefont{G.}~\bibnamefont{Cattaneo}},
  \bibinfo{journal}{Atti. Sem. Mat. Fis. Univ. Modena}
  \textbf{\bibinfo{volume}{3}}, \bibinfo{pages}{83} (\bibinfo{year}{1948}).

\bibitem[{\citenamefont{Joseph and Preziosi}(1989)}]{Joseph89}
\bibinfo{author}{\bibfnamefont{D.~D.} \bibnamefont{Joseph}} \bibnamefont{and}
  \bibinfo{author}{\bibfnamefont{L.}~\bibnamefont{Preziosi}},
  \bibinfo{journal}{Rev. Mod. Phys.} \textbf{\bibinfo{volume}{61}},
  \bibinfo{pages}{41} (\bibinfo{year}{1989}).

\bibitem[{\citenamefont{Scales and Snieder}(1999)}]{Scales99}
\bibinfo{author}{\bibfnamefont{J.~A.} \bibnamefont{Scales}} \bibnamefont{and}
  \bibinfo{author}{\bibfnamefont{R.}~\bibnamefont{Snieder}},
  \bibinfo{journal}{Nature} \textbf{\bibinfo{volume}{401}},
  \bibinfo{pages}{739} (\bibinfo{year}{1999}).

\bibitem[{\citenamefont{Valet and Fert}(1993)}]{vf}
\bibinfo{author}{\bibfnamefont{T.}~\bibnamefont{Valet}} \bibnamefont{and}
  \bibinfo{author}{\bibfnamefont{A.}~\bibnamefont{Fert}},
  \bibinfo{journal}{Phys. Rev. B} \textbf{\bibinfo{volume}{48}},
  \bibinfo{pages}{7099} (\bibinfo{year}{1993}).

\bibitem[{not()}]{note1}
\bibinfo{note}{In general, the dynamical electric field $\mathbf{E}(z,t)$ is
  related to the scalar potential $V$ and vector potential $\mathbf{A}$ by
  $\mathbf{E}=-\nabla{V}-\partial\mathbf{A}/\partial{t}$, However, if the
  magnetic field can be neglected for transport problems, one can find a gauge
  transformation that yields $E(z,t)=-\partial{V}(z,t)/\partial{z}$,
  cf.Refs.~\onlinecite{zhang02,Rash02,zhang05,cy06}.}

\bibitem[{\citenamefont{Fert and Lee}(1996)}]{fl}
\bibinfo{author}{\bibfnamefont{A.}~\bibnamefont{Fert}} \bibnamefont{and}
  \bibinfo{author}{\bibfnamefont{S.-F.} \bibnamefont{Lee}},
  \bibinfo{journal}{Phys. Rev. B} \textbf{\bibinfo{volume}{53}},
  \bibinfo{pages}{6554} (\bibinfo{year}{1996}).

\bibitem[{\citenamefont{Chester}(1963)}]{Che63}
\bibinfo{author}{\bibfnamefont{M.}~\bibnamefont{Chester}},
  \bibinfo{journal}{Phys. Rev.} \textbf{\bibinfo{volume}{131}},
  \bibinfo{pages}{2013} (\bibinfo{year}{1963}).

\bibitem[{\citenamefont{Kadin et~al.}(1980)\citenamefont{Kadin, Smith, and
  Skocpol}}]{Kadin80}
\bibinfo{author}{\bibfnamefont{A.~M.} \bibnamefont{Kadin}},
  \bibinfo{author}{\bibfnamefont{L.~N.} \bibnamefont{Smith}}, \bibnamefont{and}
  \bibinfo{author}{\bibfnamefont{W.~J.} \bibnamefont{Skocpol}},
  \bibinfo{journal}{J. Low Temp. Phys.} \textbf{\bibinfo{volume}{38}},
  \bibinfo{pages}{497} (\bibinfo{year}{1980}).

\bibitem[{\citenamefont{Ashcroft and Mermin}(1976)}]{Ash76}
\bibinfo{author}{\bibfnamefont{N.~W.} \bibnamefont{Ashcroft}} \bibnamefont{and}
  \bibinfo{author}{\bibfnamefont{N.~D.} \bibnamefont{Mermin}},
  \emph{\bibinfo{title}{Solid State Physics}}
  (\bibinfo{publisher}{Brooks/Cole}, \bibinfo{address}{Belmont, CA},
  \bibinfo{year}{1976}).

\bibitem[{\citenamefont{Yang et~al.}(1994)\citenamefont{Yang, Holody, Lee,
  Henry, Loloee, Schroeder, W.~P.~Pratt, and Bass}}]{Yang94}
\bibinfo{author}{\bibfnamefont{Q.}~\bibnamefont{Yang}},
  \bibinfo{author}{\bibfnamefont{P.}~\bibnamefont{Holody}},
  \bibinfo{author}{\bibfnamefont{S.-F.} \bibnamefont{Lee}},
  \bibinfo{author}{\bibfnamefont{L.~L.} \bibnamefont{Henry}},
  \bibinfo{author}{\bibfnamefont{R.}~\bibnamefont{Loloee}},
  \bibinfo{author}{\bibfnamefont{P.~A.} \bibnamefont{Schroeder}},
  \bibinfo{author}{\bibfnamefont{J.}~\bibnamefont{W.~P.~Pratt}},
  \bibnamefont{and} \bibinfo{author}{\bibfnamefont{J.}~\bibnamefont{Bass}},
  \bibinfo{journal}{Phys. Rev. Lett.} \textbf{\bibinfo{volume}{72}},
  \bibinfo{pages}{3274} (\bibinfo{year}{1994}).

\bibitem[{\citenamefont{Piraux et~al.}(1998)\citenamefont{Piraux, Dubois, Fert,
  and Belliard}}]{Piraux98}
\bibinfo{author}{\bibfnamefont{L.}~\bibnamefont{Piraux}},
  \bibinfo{author}{\bibfnamefont{S.}~\bibnamefont{Dubois}},
  \bibinfo{author}{\bibfnamefont{A.}~\bibnamefont{Fert}}, \bibnamefont{and}
  \bibinfo{author}{\bibfnamefont{L.}~\bibnamefont{Belliard}},
  \bibinfo{journal}{Eur. Phys. J. B.} \textbf{\bibinfo{volume}{4}},
  \bibinfo{pages}{413} (\bibinfo{year}{1998}).

\bibitem[{\citenamefont{Hershfield and Zhao}(1997)}]{Her97}
\bibinfo{author}{\bibfnamefont{S.}~\bibnamefont{Hershfield}} \bibnamefont{and}
  \bibinfo{author}{\bibfnamefont{H.~L.} \bibnamefont{Zhao}},
  \bibinfo{journal}{Phys. Rev. B} \textbf{\bibinfo{volume}{56}},
  \bibinfo{pages}{3296} (\bibinfo{year}{1997}).

\bibitem[{\citenamefont{Masoliver and Weiss}(1996)}]{Mas96}
\bibinfo{author}{\bibfnamefont{J.}~\bibnamefont{Masoliver}} \bibnamefont{and}
  \bibinfo{author}{\bibfnamefont{G.~H.} \bibnamefont{Weiss}},
  \bibinfo{journal}{Eur. J. Phys.} \textbf{\bibinfo{volume}{17}},
  \bibinfo{pages}{190} (\bibinfo{year}{1996}).

\bibitem[{\citenamefont{Ames}(1992)}]{Ames92}
\bibinfo{author}{\bibfnamefont{W.~F.} \bibnamefont{Ames}},
  \emph{\bibinfo{title}{Numerical Methods for Partial Differential Equations}}
  (\bibinfo{publisher}{Academic Press, San Diego, California},
  \bibinfo{year}{1992}), \bibinfo{edition}{3rd} ed., \bibinfo{note}{see Sec.
  4-15}.

\end{thebibliography}

\end{document}